\documentclass[journal=jpcbfk,articletitle=true,manuscript=article]{achemso}
  
\usepackage{graphicx}
\usepackage{amsmath}
\usepackage{amssymb}
\usepackage{dcolumn}
\usepackage{color}
\usepackage{natbib}
\usepackage{ctable}
\usepackage[sort&compress]{natbib}

\title{$^1$H-NMR Dipole-Dipole Relaxation in Fluids: Relaxation of Individual $^1$H-$^1$H Pairs versus Relaxation of Molecular Modes}
\author{D.\ Asthagiri}
\email{dna6@rice.edu}
\affiliation{Rice University, Department of Chemical and Biomolecular Engineering, 6100 Main St., Houston, TX 77005, USA}
\author{Walter G.\ Chapman}
\affiliation{Rice University, Department of Chemical and Biomolecular Engineering, 6100 Main St., Houston, TX 77005, USA}
\author{George J.\ Hirasaki}
\affiliation{Rice University, Department of Chemical and Biomolecular Engineering, 6100 Main St., Houston, TX 77005, USA}
\author{Philip M.\ Singer}
\email{ps41@rice.edu}
\affiliation{Rice University, Department of Chemical and Biomolecular Engineering, 6100 Main St., Houston, TX 77005, USA}

\begin{document}

\begin{abstract}
The intra-molecular $^1$H-NMR dipole-dipole relaxation of molecular fluids has traditionally been interpreted within the Bloembergen-Purcell-Pound (BPP) theory of NMR intra-molecular relaxation. The BPP theory draws upon Debye's theory for describing the rotational diffusion of the $^1$H-$^1$H pair and predicts a mono-exponential decay of the $^1$H-$^1$H dipole-dipole autocorrelation function between distinct spin pairs. Using molecular dynamics (MD) simulations, we show that for both $n$-heptane and water this is not the case. In particular, the autocorrelation function of individual $^1$H-$^1$H intra-molecular pairs itself evinces a rich stretched-exponential behavior, implying a distribution in rotational correlation times. However for the high-symmetry molecule neopentane, the individual $^1$H-$^1$H intra-molecular pairs do conform to the BPP description, suggesting an important role of molecular symmetry in aiding agreement with the BPP model. The inter-molecular autocorrelation functions for $n$-heptane, water, and neopentane also do not admit a mono-exponential behavior of individual $^1$H-$^1$H inter-molecular pairs at distinct initial separations. 
We suggest expanding the auto-correlation function in terms of molecular modes, where the molecular modes do have an exponential relaxation behavior. With care, the resulting Fredholm integral equation of the first kind can be inverted to recover the probability distribution of the molecular modes. The advantages and limitations of this approach are noted. 
\end{abstract}

\begin{tocentry}
\begin{center}
        \includegraphics[scale=1.0]{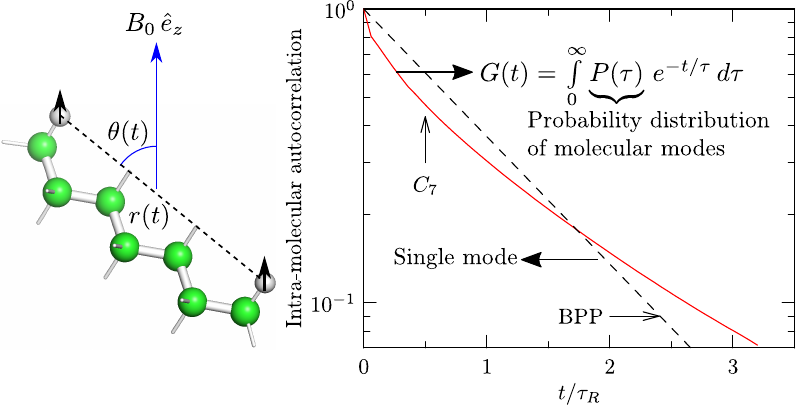}
\end{center}
\end{tocentry}

\newpage

\section{Introduction}\label{sc:Intro}

In nuclear magnetic resonance (NMR) relaxation experiments, the nuclear magnetic moments (i.e. the nuclear spins) in the sample are aligned using a static magnetic field and then suitably perturbed using an oscillating magnetic field perpendicular to the static field. The relaxation time back to equilibrium following the perturbation is interpreted to gain insights into fluctuations of the local magnetic fields. 

In fluids, the fluctuations in local magnetic fields happen primarily due to two effects: (1) the tumbling of the nuclear spins on the same molecule, which is responsible for intra-molecular relaxation, and (2) the relative motion between nuclear spins on different molecules, which is responsible for inter-molecular relaxation. Intra-molecular relaxation is ascribed to rotational diffusion, while inter-molecular relaxation is ascribed to translational diffusion. In the case of liquid $n$-alkanes and water, molecular dynamics (MD) simulations previously showed that intra-molecular relaxation dominates, especially with increasing carbon number \cite{singer:jmr2017,asthagiri:seg2018}. On the other hand, for benzene and cyclohexane, inter-molecular relaxation dominates \cite{singer:jcp2018}. In other words, the molecular geometry and internal motions play a crucial role in the origins of the NMR relaxation mechanism in fluids.

Bloembergen, Purcell, and Pound (BPP) pioneered the use of NMR and specifically considered intra-molecular relaxation for a pair of nuclear spins \cite{bloembergen:pr1948}. Treating each spin pair as a rotationally diffusing unit, BPP assumed an exponential decay with time (similar to the Debye model) of the intra-molecular autocorrelation between the spin pair. Using MD simulations of a series of $n$-alkanes and water, we previously showed that the decay of the intra-molecular autocorrelation for $^1$H spin pairs on the same molecule does not conform to a mono-exponential behavior \cite{singer:jmr2017,singer:jcp2018,asthagiri:seg2018}. It is important to emphasize that in that study, even for water, which has only a pair of spins, the autocorrelation did not conform to a mono-exponential behavior. But water is also unique because of the strong inter-molecular hydrogen bonding and it is possible that this invalidates the freely rotating picture \cite{madhavi:jpcb2017}. 

More recently, researchers sought to understand NMR relaxation in a molecular dynamics simulation of an ionic liquid \cite{hoenegger:jpc2020}. These researchers suggested that the decay of the intra-molecular autocorrelation of $^1$H spin pairs on the same molecule can be fit to a mono-exponential decay for distinct spin pairs. The researchers also described the inter-molecular autocorrelation of $^1$H spin pairs on different molecules can be fit to a mono-exponential decay for distinct spin pairs and distinct initial spin-pair separations. These results stand in sharp contrast to our earlier study, and raise the natural question whether the behavior that we found for the intra-molecular relaxation in $n$-alkanes can be described by treating distinct spin pairs as distinct rotationally diffusing units. 

Here we study the aforementioned question for $n$-heptane, neopentane, and water. We reason that if a mono-exponential decay of the autocorrelation function for distinct $^1$H spin pairs holds for an ionic liquid, it must also hold for $n$-heptane and neopentane, fluids that admit a van~der~Waals description. However, we find that the intra-molecular relaxation in water does not conform to a mono-exponential behavior, and importantly, the relaxation of distinct intra-molecular autocorrelation function for spin pairs of $n$-heptane also does not evince a mono-exponential behavior. Likewise, for all the fluids, the inter-molecular autocorrelation function of distinct spins pairs and distinct initial separation do not admit a mono-exponential behavior. Overall, our results show that much care is needed in adapting ideas from the traditional BPP theory to predict the NMR relaxation dispersion (i.e. frequency dependence) from MD simulations. 

In light of these findings, we propose a general solution to interpret the autocorrelation function. We propose expanding the autocorrelation in terms of molecular modes, where each mode admits an exponential relaxation behavior. The resulting Fredholm integral equation of the first kind can be inverted with care to uncover the probability distribution of the modes. For concision and following precedence in the literature, we call this ``inverse Laplace transform" (ILT) \cite{singer:jcp2018,singer:jcp2018b,parambathu:jpcb2020,singer:jpcb2020}. The ILT analysis does not depend on a  model for the autocorrelation function of spin-pairs such as the mono-exponential decay by BPP or stretched exponents \cite{singer:prb2020}. Instead, the ILT of the autocorrelation function yields a distribution of correlation times, which can then be used to predict the NMR relaxation dispersion, without assuming a model of the molecular motion.   The resulting NMR relaxation dispersion from ILT analysis was previously shown to agree with measurements in the case of viscous polymers \cite{singer:jpcb2020} and $n$-heptane under nano-confinement in a polymer matrix \cite{parambathu:jpcb2020}, which validates the approach. 

\section{Methods}

\subsection{Autocorrelation of spin magnetic moments} 

Following our earlier studies, for an {\it isotropic} system, the autocorrelation function $G(t)$ for fluctuating magnetic $^1$H-$^1$H dipole-dipole interactions is given as \cite{mcconnell:book,cowan:book}: 
\begin{eqnarray}
G_{R,T}(t) = \frac{3}{16} \! \left(\frac{\mu_0}{4\pi}\right)^2 \! \hbar^2 \gamma^4  \frac{1}{N_{R,T}} \! \sum\limits_{i \neq j}^{N_{R,T}} \! \left\langle \frac{(3\cos^{2}\!\theta_{ij}\!(t+\tau)-1)}{r_{ij}^3\!\left(t+\tau\right)}  \frac{(3\cos^{2}\!\theta_{ij}\!(\tau)-1)}{r_{ij}^3\!(\tau)} \right\rangle_{\!\! \tau} \label{eq:auto}
\end{eqnarray}
where $t$ is the lag time of the autocorrelation, $\tau$ is the trajectory time in the simulation, $\mu_0$ is the vacuum permeability, $\hbar$ is the reduced Planck constant, $\gamma/2\pi = 42.58$ MHz/T is the nuclear gyro-magnetic ratio for $^1$H (spin $I=1/2$), $r_{ij}$ is the magnitude of the vector connecting the $(i,j)$ $^1$H-$^1$H dipole-pairs, and $\theta_{ij}$ is the polar angle 
between $\vec{r}_{ij}$ and the external magnetic field. The subscript $R$ refers to intra-molecular interactions from rotational diffusion, while the subscript $T$ refers to inter-molecular interactions from translational diffusion. 

An equivalent form of Eq. \ref{eq:auto} for an {\it isotropic} system is \cite{cowan:book}:
\begin{eqnarray}
G_{R,T}(t) & = & \frac{3}{16} \! \left(\frac{\mu_0}{4\pi}\right)^2 \! \hbar^2 \gamma^4 \frac{1}{N_{R,T}} \! \sum\limits_{i \neq j}^{N_{R,T}} \! \frac{2}{5} \! \left\langle \frac{(3\cos^{2}\!\psi_{ij}(t+\tau)-1)}{r_{ij}^3\!\left(t+\tau\right) r_{ij}^3\!(\tau)}  
\right\rangle_{\!\! \tau} \label{eq:auto2}
\end{eqnarray}
where $\psi_{ij}(t+\tau)$ is the angle between $\vec{r}_{ij}\!\left(t+\tau\right)$ and $\vec{r}_{ij}\!\left(\tau\right)$. Eqs. \ref{eq:auto} and \ref{eq:auto2} predict a quantitative value for $G_{R,T}(t)$, and therefore a quantitative value for the NMR relaxation times, {\it without any adjustable parameters} \cite{singer:jmr2017}. This is an important step for validating the MD simulations against NMR measurements, which does not rely on adjustable parameters. 

The NMR spectral density function $J_{R,T}(\omega)$ is determined from the Fourier transform of $G_{R,T}(t)$ as such:
\begin{equation}
J_{R,T}(\omega) = 2\int_{0}^{\infty}G_{R,T}(t)\cos\left(\omega t\right) dt,
\label{eq:FourierRTcos}
\end{equation}
where $G_{R,T}(t)$ (in units of $\rm{s}^{-2}$) is real and an even function of $t$. $J_{R,T}(\omega)$ is then used to compute $T_1$ and $T_2$ as a function of the Larmor frequency $\omega_0 = 2 \pi f_0$, i.e.\ the $T_1$ and $T_2$ dispersion, using the following expressions (which do not assume a molecular model) \cite{singer:jmr2017}:
\begin{align}
\frac{1}{T_{1R,1T}} &= J_{R,T}(\omega_0) + 4 J_{R,T}(2\omega_0), \label{eq:T1RT} \\
\frac{1}{T_{2R,2T}} &= \frac{3}{2} J_{R,T}(0) + \frac{5}{2} J_{R,T}(\omega_0) + J_{R,T}(2\omega_0), \label{eq:T2RT}\\
\frac{1}{T_{1,2}} &= \frac{1}{T_{1R,2R}} + \frac{1}{T_{1T,2T}}. \label{eq:T12}
\end{align}
Note that the intra-molecular and inter-molecular rates add to give the total relaxation rate (Eq. \ref{eq:T12}).

One limitation with the MD simulations is that the typical maximum autocorrelation time computed for $G_{R,T}(t)$ is $t_{max} \simeq $ 1 ns, which is limited by computational cost. Assuming $t_{max} = $ 1 ns, the bin width (i.e. resolution) of $J_{R,T}(\omega)$ from Eq. \ref{eq:FourierRTcos} is $\Delta f = 1/2t_{max} = $ 500 MHz. In other words, using Eq. \ref{eq:FourierRTcos} on the $G_{R,T}(t)$ data directly cannot determine dispersion below $f_0 < 500$ MHz, which is much larger than typical dispersion results require. 

A work around for predicting $J_{R,T}(\omega)$ below $f_0 \lesssim 500$ MHz is to assume a model for $G_{R,T}(t)$ above $t > t_{max} $. One such model is the BPP picture used in the recent study of ionic liquids \cite{hoenegger:jpc2020}, where the autocorrelation function of a rotating spin-pair $k = ij$ obeys:
\begin{eqnarray}
G_{R,k}(t) = A_k \exp\left(-\frac{t}{\tau_{R,k}}\right) \,, \label{eq:Gt}
\end{eqnarray}
where $A_k \propto 1/r^6_k$. The summation over all spin pairs $k$ is then given by: 
\begin{eqnarray}
G_R(t) = \sum_k^{\{pairs\}} A_k  \exp\left(-\frac{t}{\tau_{R,k}}\right) \, , \label{eq:SumExp}
\end{eqnarray}
which assumes that each distinct pair of intra-molecular spins can be fit to a mono-exponential, and where the proportionality constant is a free parameter. As shown below, we show that using Eq. \ref{eq:SumExp} for spin-pairs is not accurate in the relatively simple cases of $n$-heptane and water. Consequently, it is logical that using Eq. \ref{eq:SumExp} for more complex fluids such as ionic liquids is not accurate, implying that the resulting $T_1$ and $T_2$ dispersion will also not be accurate.

\subsection{Simulation Details}

We follow our earlier study \cite{singer:jmr2017} in modeling the system. The molecular simulations were performed using NAMD \cite{namd1} version 2.11. The bulk alkanes were modeled using the CHARMM General Force field, CGenFF \cite{cgenff:2010}. Water was described using the TIP4P/2005 model \cite{tip4p}. 

For $n$-heptane, we used the data from our earlier study\cite{singer:jmr2017}. In that study, the system was rigorously equilibrated at 20 $^\circ$C by reassigning velocities (obtained from a Maxwell-Boltzmann distribution) every 250 fs. Subsequently, the production run lasted  2~ns under $NVE$ conditions. The time step for integration was 1~fs, and during the production phase configurations were archived every 100~fs for analysis. We used the last 16384 ($=2^{14}$) frames out of the total 20,000 frames for autocorrelation analysis. 

For TIP4P/2005 water simulation, because we did not have the final velocities from our earlier study  \cite{singer:jmr2017}, the final configuration from our earlier $NVE$ simulations was once again equilibrated under $NVT$ conditions for over 0.5~ns.  Subsequently, the production phase was in the $NVE$ ensemble with frames archived every 100~fs. The average temperature in the NVE phase was 296~K. (We use SHAKE \cite{md:shake} to constrain the structure of water.) The production phase lasted 2~ns with frames archived every 100~fs. 

For neopentane, as for TIP4P/2005, we took the final configuration from our earlier study \cite{singer:jcp2018} and after equilibrating under $NVT$ for 1~ns, we ran the production under $NVE$ conditions. The production phase lasted 2~ns with frames archived every 100~fs. The average temperature in the NVE phase was 294~K. 

In all the simulations, the Lennard-Jones interactions were terminated at 14.00 {\AA} (11 {\AA} for water) by smoothly switching to zero starting at 13.00 {\AA} (10 {\AA} for water). Electrostatic interactions were treated with the particle mesh Ewald method with a grid spacing of 0.5 {\AA}; the real-space contributions to the electrostatic interaction were cutoff at 14.00 {\AA}.  As before \cite{singer:jmr2017}, the autocorrelation function $G_{R,T}(t)$ was constructed using fast Fourier transforms, for lag time ranging from $0$ ps to $\approx$75~ps in steps of $0.1$ ps.

\section{Results and Discussion}

\subsection{Intra-molecular relaxation}
\begin{figure}[ht!]
\includegraphics[scale=0.9]{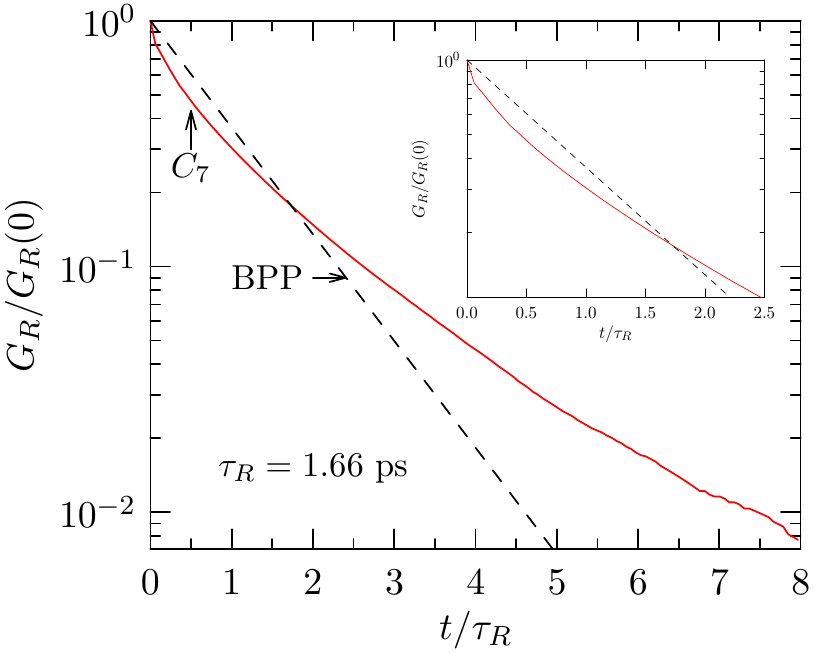} \\
\includegraphics[scale=0.9]{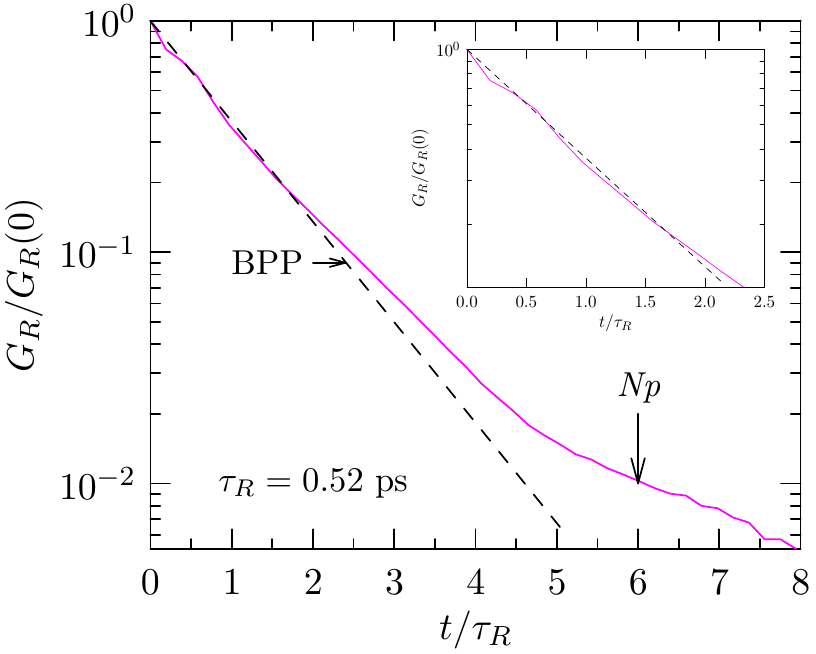} \\
\includegraphics[scale=0.9]{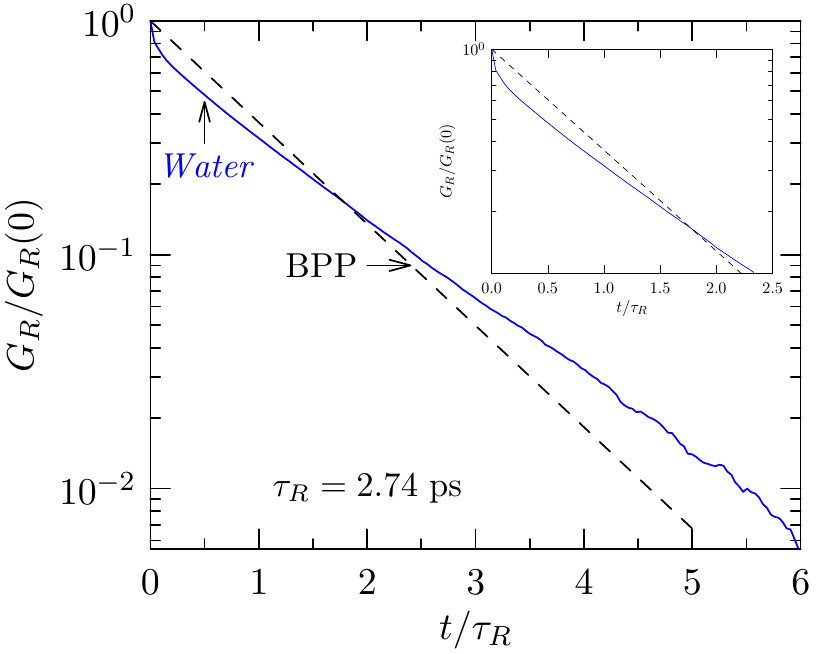}
\caption{Normalized autocorrelation of all the intra-molecular spin-pairs summed together. The mono-exponential decay model is the one due to the BPP theory. The correlation time (Eq.~\ref{eq:autotime}) is noted in each figure. The inset shows the short time behavior.}\label{fg:Intrabulk}
\end{figure}

Figure~\ref{fg:Intrabulk} shows the autocorrelation of intra-molecular spin interactions as the molecule undergoes rotational diffusion. 
To better compare different fluids, we normalize the $x$-axis by the correlation time $\tau_{R}$, where
\begin{eqnarray}
\tau_{R} = \frac{1}{G_R(0)} \int\limits_0^\infty G_R(t)\, dt \, , 
\label{eq:autotime}
\end{eqnarray}
and we normalize the $y$-axis by $G_R(0)$.  Figure~\ref{fg:Intrabulk} makes it clear that $G_R(t)$ cannot be described by a mono-exponential decay. In fact for water, which has only a single pair of spins, $G_R(t)$ is clearly not mono-exponential, in contrast to what BPP assumed \cite{bloembergen:pr1948}. Among the molecules considered, the tetra symmetric molecule neopentane comes closest to being spherical, and therefore closest to obeying BPP. 

We next consider whether the observed decay (Figure~\ref{fg:Intrabulk}) for $n$-heptane or neopentane can be modeled by a sum over the decay of pairs of intra-molecular spins, with each pair assumed to relax in an exponential fashion (Eq.~\ref{eq:SumExp}).  To this end, using linear regression we fit the first 1~ps of $G_R(t)$ of each spin pair to the function 
\begin{eqnarray}
G_R(t) = G_R(0) \exp\left(- \frac{t} {t^\star}\right) \label{eq:tauStar}
\end{eqnarray}
where $G_R(0)$ and $t^\star$ are free parameters. Subsequently, for the fit function we shift the intercept to 0. If the decay does conform to a mono-exponential behavior, then the slope $(-1/t^\star)$ should agree with the slope of the normalized autocorrelation ($G_R(t)/ G_R(0)$) obtained from simulations. Please note that we use $\tau^\star$ to differentiate the fit time-constant from the 
bona fide autocorrelation time obtained using Eq.~\ref{eq:autotime}. Also, as would be clear below, our physical conclusions are insensitive to the choice of 1~ps length of data used in the fitting procedure.

\begin{figure}[h!]
\includegraphics[scale=0.9]{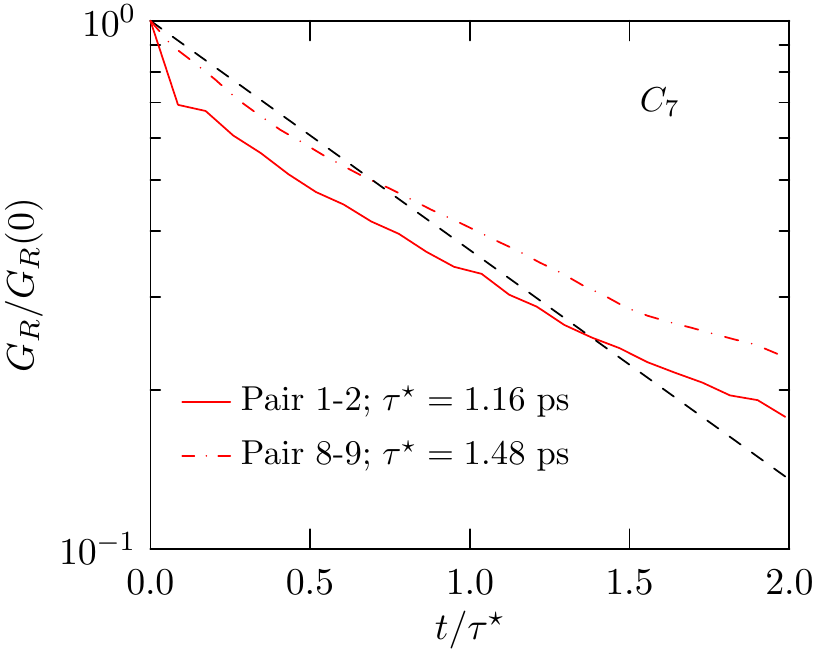} \\
\includegraphics[scale=0.9]{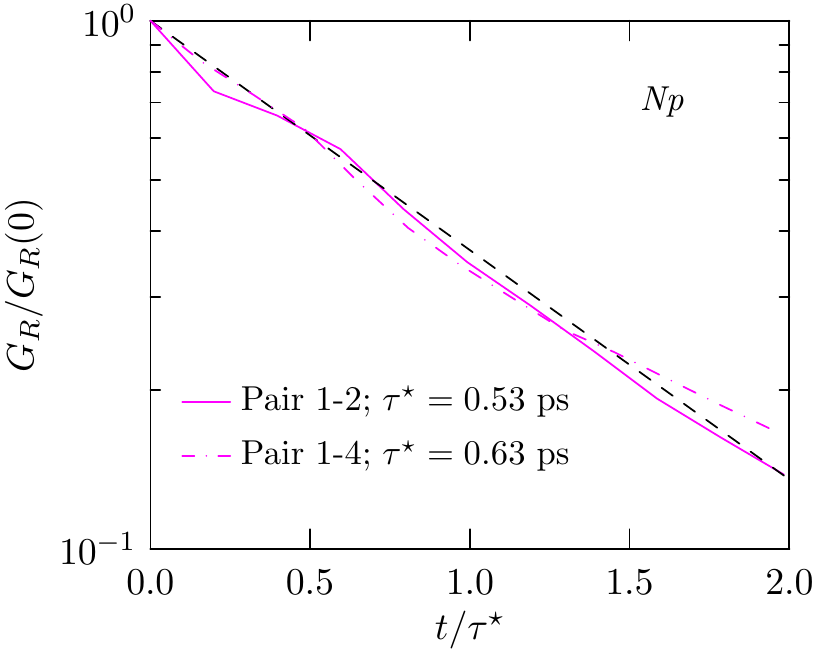} \\
\includegraphics[scale=0.9]{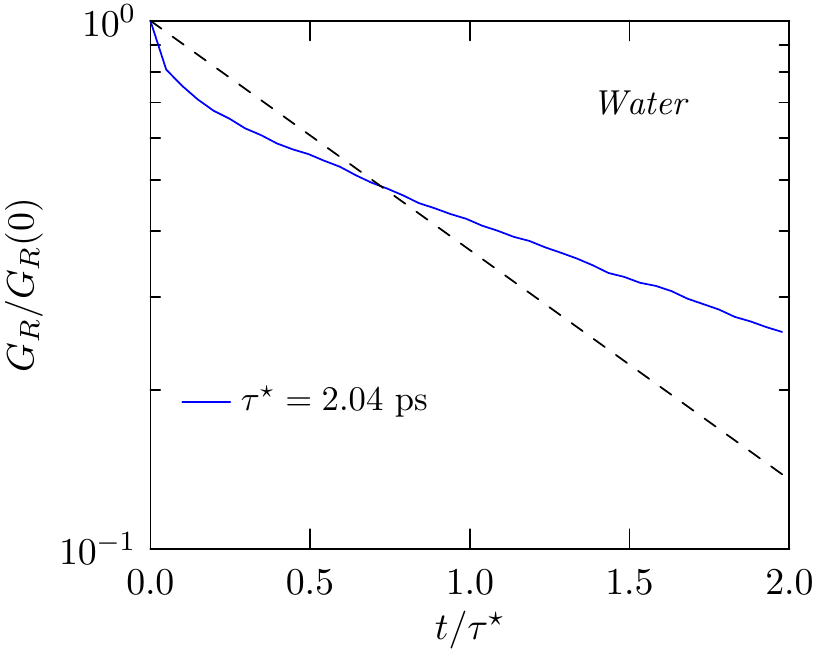}
\caption{Normalized autocorrelation function of distinct pairs of intra-molecular spins. ``Pair $m$-$n$" refers to the pair formed by protons $m$ and $n$. For $C_7$, the $^1$H nuclei are labeled according to the following scheme: $C_{{\rm H}_1, {\rm H}_2, {\rm H}_3}-C_{{\rm H}_4,{\rm H}_5}-C_{{\rm H}_6,{\rm H}_7}-C_{{\rm H}_8,{\rm H}_9}-C_{{\rm H}_{10},{\rm H}_{11}}-C_{{\rm H}_{12},{\rm H}_{13}}-C_{{\rm H}_{14},{\rm H}_{15},{\rm H}_{16}}$. For neopentane, we follow a similar scheme. For water, there is  only one intra-molecular pair. The mono-exponential behavior is shown by the dashed line. The correlation time (Eq.~\ref{eq:tauStar}) is noted in each figure.}\label{fg:intrapairs}
\end{figure}
$n$-Heptane has 16 protons and thus there are 120 distinct pairs of protons and for neopentane we need to consider 
66 distinct spin pairs. But to test whether each pair conforms to the mono-exponential behavior, it proves helpful to focus on only a select subset of this rather large set of pairs. To this end, we select particular molecules from the simulation trajectory and for the particular molecule, extract a defined spin pair and compute the autocorrelation. (Please note this procedure conforms to that
suggested in Ref.~\citenum{hoenegger:jpc2020}.)  Figure~\ref{fg:intrapairs} shows the autocorrelation functions for distinct pairs of intra-molecular spins.  It is obvious that for $n$-heptane and water, the autocorrelation for distinct spin-pairs does not admit a mono-exponential behavior.  However, for neopentane, the autocorrelation does closely conform to the BPP model.  This makes good physical sense since neopentane is a high-symmetry molecule, and also expected to be fairly rigid since the carbon-hydrogen bond length and bond angle fluctuations are not expected to be large. Thus the tumbling of intra-molecular spin-pairs in neopentane is expected to better conform to the assumptions underlying the BPP theory. 

The above result brings us to an important conclusion. In general, the relaxation of pairs of intra-molecular $^1$H spins will not conform to the BPP description, except in cases involving fairly rigid molecules with a high degree of symmetry. 
We can thus safely conclude that if Eq.~\ref{eq:SumExp} cannot form an adequate basis to model the pair-wise intra-molecular auto-correlation for simple cases such as $n$-heptane and water (Fig.~\ref{fg:intrapairs}), it will fail for more complex cases such as ionic liquids \cite{hoenegger:jpc2020}. As such, Eq.~\ref{eq:SumExp} cannot reliably predict the dispersion of intra-molecular NMR relaxation times for simple fluids, let alone ionic liquids.
 
\subsection{Inter-molecular relaxation}
 
\begin{figure}[h!]
\includegraphics[scale=0.85]{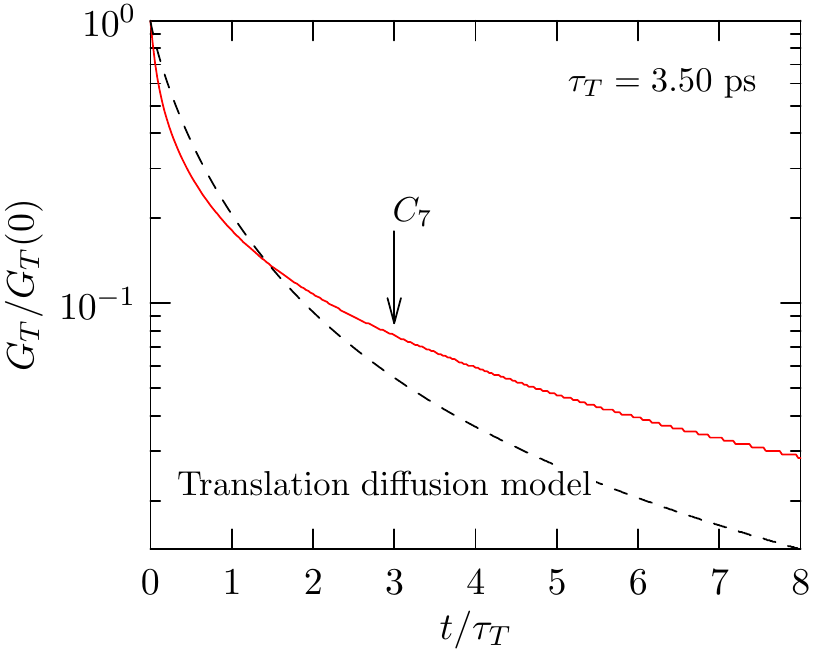} \\
\includegraphics[scale=0.85]{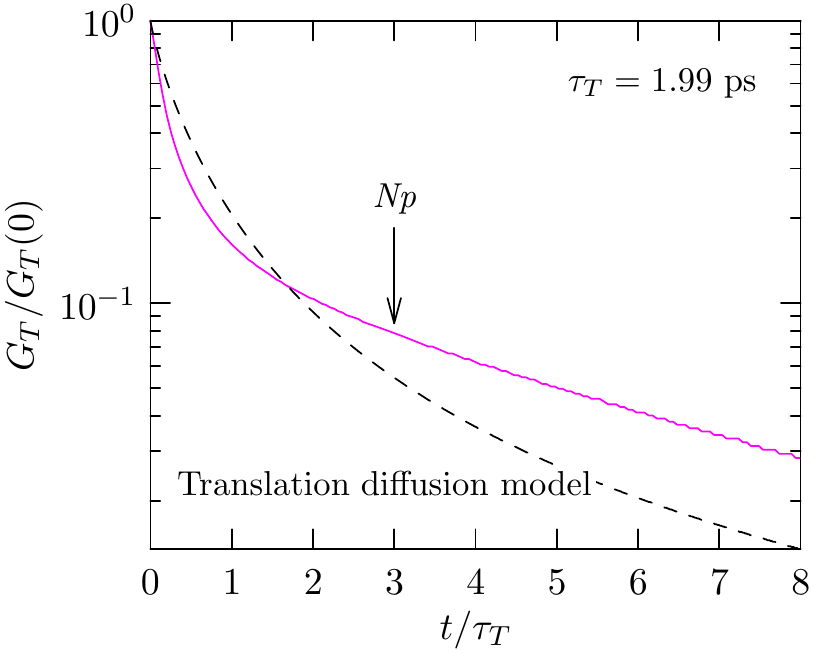} \\
\includegraphics[scale=0.85]{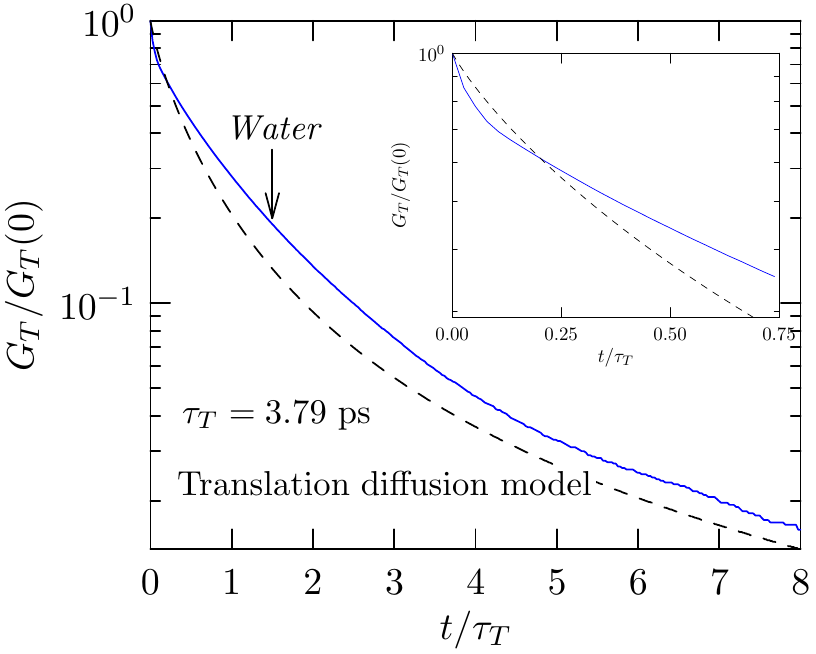}
\caption{Normalized autocorrelation of all inter-molecular spin-pairs summed together. For water, the inset highlights the deviation from the model at short times. The translational diffusion model due to Hwang and Freed \cite{hwang:JCP1975} is shown by a dashed line. Rest as in Figure~\ref{fg:Intrabulk}. }\label{fg:InterBulk}
\end{figure}

For a hard-sphere fluid, by building on a previous theory by Torrey \cite{torrey:pr1953} and explicitly incorporating finite size effects (i.e. a distance of minimum approach) into relaxation theory, Hwang and Freed \cite{hwang:JCP1975} have shown that the autocorrelation due to translational diffusion should obey:
\begin{eqnarray}
G_T(t) \propto \frac{54}{\pi}\int\limits_0^\infty  \frac{x^2}{81 + 9 x^2 - 2 x^4 + x^6} \exp\left(-x^2 \frac{t}{\frac{9}{4} \tau_T} \right)  dx \label{eq:hwang}
\end{eqnarray}
where the factor $9/4$ is explained in Ref.~\citenum{singer:jcp2018}. Figure~\ref{fg:InterBulk} shows the inter-molecular autocorrelation by translational diffusion for the molecules considered in this work.  The numerical agreement with the hard-sphere model is necessarily deficient, but qualitatively one can observe a similar decay between the model and the actual data. Importantly, it is evident that the decay does not conform to a mono-exponential decay. 

\begin{figure}[h!]
\includegraphics[scale=0.85]{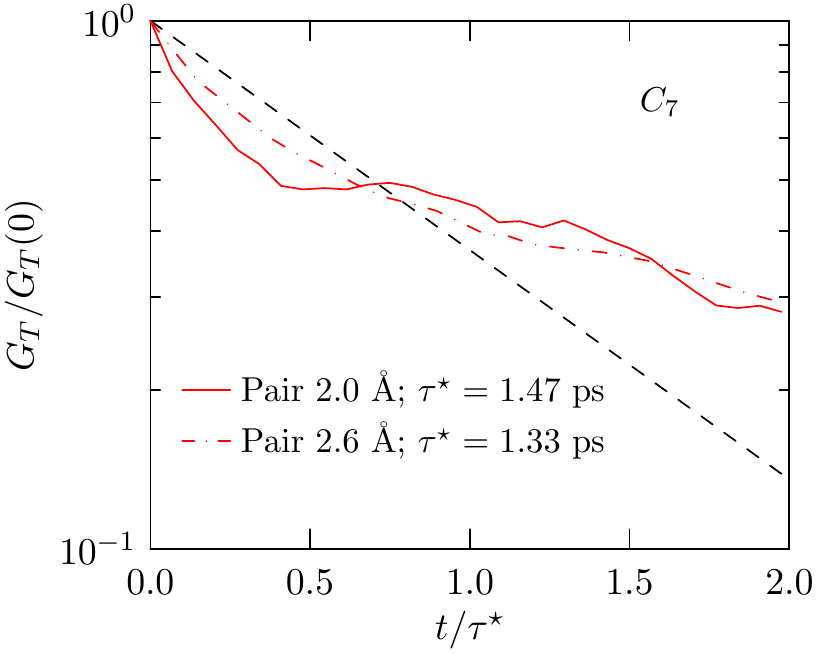}\\
\includegraphics[scale=0.85]{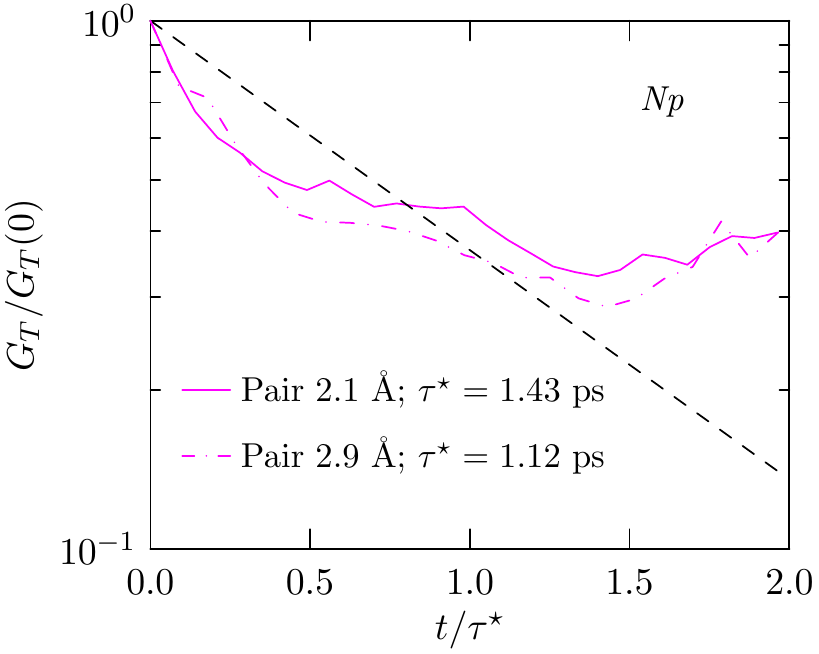}\\
\includegraphics[scale=0.85]{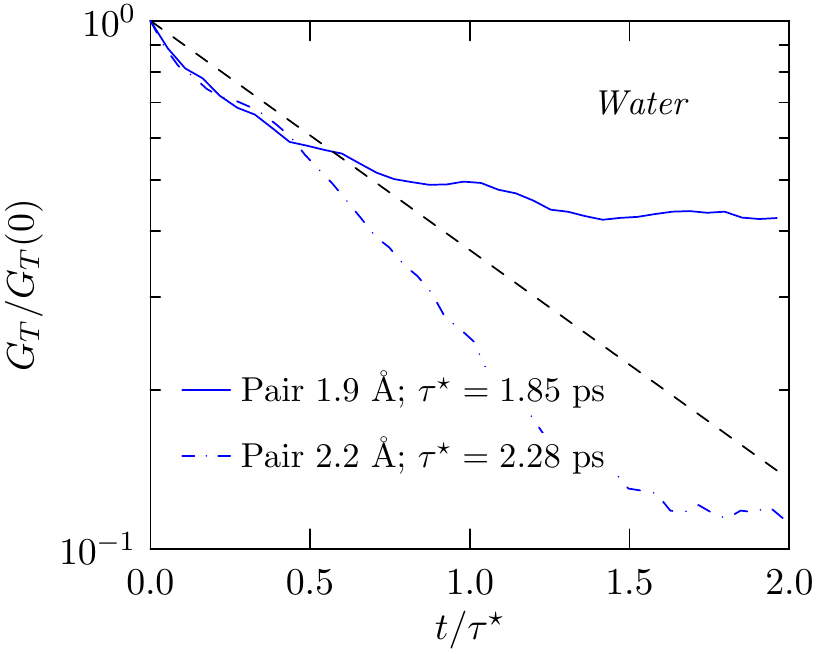}\\
\caption{Normalized autocorrelation of inter-molecular spin-pair interactions for select spin-pairs in $n$-heptane, neopentane, and water. The initial separation of the spin-pair is noted in the legend. The dashed line is a mono-exponential fit to the first 1~ps of the data. The fit function is then translated to have an intercept of 0 (on the log-scale).The correlation time (Eq.~\ref{eq:tauStar}) is noted in each figure.}\label{fg:interpair}
 \end{figure}
Following the claims by the recent study \cite{hoenegger:jpc2020}, we asked if the autocorrelation function of spin-pairs on different molecules can be adequately described by a mono-exponential decay. Figure~\ref{fg:interpair} compares the autocorrelation of select spin-pairs, whose initial separations are noted in the figure.  As done in Fig.~\ref{fg:intrapairs}, 
we fit Eq. \ref{eq:tauStar} to the first 1~ps of the data and then shift the intercept to 0. Clearly, the inter-molecular autocorrelation for select spin pairs does not conform to a mono-exponential behavior, contrary to what has been suggested recently \cite{hoenegger:jpc2020}.

Note that in contrast to intra-molecular spin pairs (Fig.~\ref{fg:intrapairs}), for inter-molecular spin pairs we see an
increase in the noise as the lag time increases. Note that the noise takes the form of oscillations rather than Gaussian noise, although a 2$\sigma$ uncertainty can still be computed (see Supplementary Material in Ref. \citenum{singer:jcp2018}). 

To better understand the origin of noise, we constructed synthetic data-sets for the diffusive evolution of the distance, $r$, between a pair of spins (data not shown). The angle $\theta$ (Eq.~\ref{eq:auto}) is also held fixed. From such synthetic data sets, we find that the oscillations occur when the reaction coordinate, $r$, makes periodic returns to smaller $r$ values and dwells around that value before escaping to a different  value.  The effect of such behavior is expected to be washed out when we average over many different pairs and also allow for rotation of the vector connecting the spins, as we do in computing the overall autocorrelation (Fig.~\ref{fg:InterBulk}).  Nevertheless, Figure~\ref{fg:interpair} makes it clear that the decay of autocorrelation function for select spin-pairs with distinct initial separations does not conform to a mono-exponential decay, emphasizing that it is incorrect to model the overall decay curve using Eq.~\ref{eq:SumExp}. As such, Eq.~\ref{eq:SumExp} cannot accurately predict the dispersion (i.e. the frequency dependence) of inter-molecular NMR relaxation times.

\subsection{Proposed solution}
 
In order to surmount the computational limitations of maximum autocorrelation time $t_{max} $ in $G_{R,T}(t)$, and thereby surmount the limitations in determining $T_1$ and $T_2$ dispersion (especially at frequencies below $f_0 \lesssim 500$ MHz), one has to make predictions of $G_{R,T}(t)$ above $t > t_{max} $. We have shown that using Eq.~\ref{eq:SumExp} to model $G_{R,T}(t)$ is not accurate for spin-pairs on relatively simple molecules such as $n$-heptane, and water. It therefore follows that using Eq. \ref{eq:SumExp} will be even more inaccurate for more complex fluids such as ionic liquids \cite{hoenegger:jpc2020}, thereby leading to inaccuracies in predicting the $T_1$ and $T_2$ dispersion. We also note that there are no analytic expressions or theories to extend the $G_{R,T}(t)$ beyond $t > t_{max} $.

One solution we have developed is to expand $G_{R,T}$ as 
\cite{venkataramanan:ieee2002,song:jmr2002}: 
\begin{align}
G_{R,T}(t) &= \int_{0}^{\infty}\! P_{R,T}(\tau) \exp\left(-\frac{t}{\tau}\right) d\tau, \label{eq:ILT1}
\end{align}
where $P_{R,T}(\tau)$ is the distribution in molecular correlation times. Our aim is to recover $P_{R,T}$ from Eq.~\ref{eq:ILT1}, which is a Fredholm integral equation of the first kind. For concision, and because this terminology is widely used, we term the procedure of recovering $P_{R,T}$ an ``inverse Laplace transform,"(ILT) but we emphasize that inverting Eq.~\ref{eq:ILT1} to recover $P_{R,T}$ is not formally a Laplace inversion \cite{fordham:diff2017}. With this understanding, we note that details of the ILT procedure can be found in Refs.\ \citenum{parambathu:jpcb2020,singer:jpcb2020,singer:jcp2018b} and the supplementary material in Ref.~\citenum{singer:jcp2018,singer:prb2020}. We briefly highlight the essential details here. 

At the outset please note that $G_{R,T}$ is available only at discrete time intervals, and moreover, the complete $G_{R,T}$ is also not available, for we are limited by the longest times we can simulate. Thus the inversion is a rather ill-posed problem. In our approach, we use Tikhonov regularization \cite{singer:jcp2018,singer:prb2020}, with the vector $\textbf{P}$ being one for which 
\begin{eqnarray}
|| \textbf{G}- K \textbf{P} ||^2 + \alpha ||\textbf{P}||^2
\label{eq:reg}
\end{eqnarray}
is a minimum. Here $\textbf{G}$ is the column vector representation of the autocorrelation function $G_{R,T}(t)$, $\textbf{P}$ is the column vector representation of the distribution function $P_{R,T}(\tau)$, $\alpha$ is the regularization parameter, and $K$ is the kernel matrix:
\begin{align}
K = K_{ij} = \exp\left(-\frac{t_i}{\tau_{j}}\right). \label{eq:kernel}
\end{align}

In the cases we have studied so far, the inversion is found to be well-determined. Please note that for this problem of inversion other methods also exist, such as maximum entropy reconstruction \cite{sibisi:jmr85,sibisi:nature83,maxentILT:2010}, using both $L_1$ and $L_2$ norms \cite{wiesman:ilt13} in Eq.~\ref{eq:reg}, or describing the Laplace transform of $G_{R,T}$ using Pad{\'e} approximants \cite{pade:nature87}. It may also be possible to combine ideas from these different methods. Indeed exploiting maximum entropy modeling, an idea that has proven very successful in reconstructing probability distributions with sparse data, is part of our on-going investigations. 

Once $P_{R,T}(\tau)$ are determined from Eq. \ref{eq:ILT1}, the spectral density $J_{R,T}(\omega)$ is straightforwardly determined from the Fourier transform (Eq. \ref{eq:FourierRTcos}) of $G_{R,T}(t)$ (Eq. \ref{eq:ILT1}): \begin{align}
J_{R,T}(\omega) &= \int_{0}^{\infty}\! \frac{2\tau}{1+(\omega\tau)^2}   P_{R,T}(\tau) d\tau \, , \label{eq:JwfromPt}
\end{align}
from which $T_1$ and $T_2$ at $\omega_0$ can be determined \cite{singer:jmr2017}.

\begin{figure}[h!]
\includegraphics[scale=0.9]{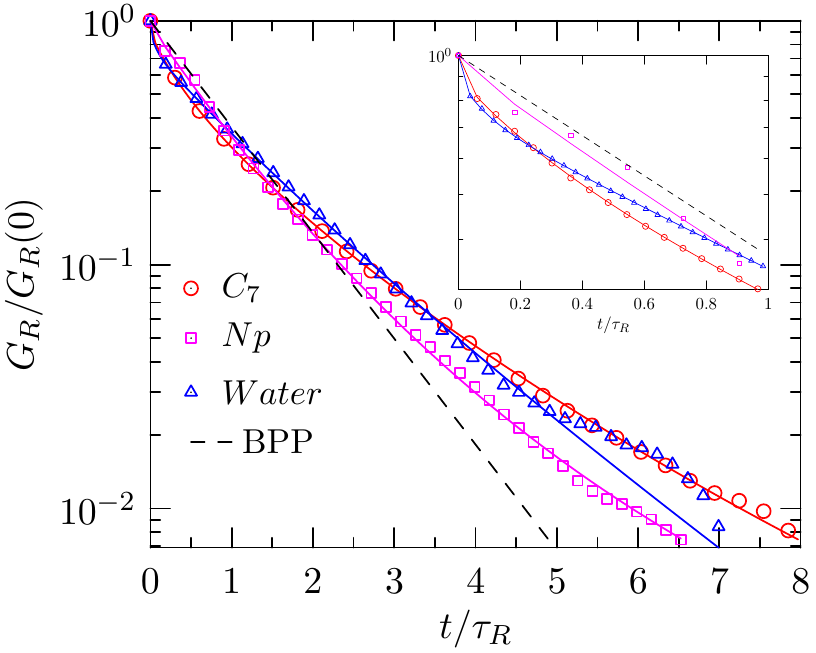}\\
\includegraphics[scale=0.9]{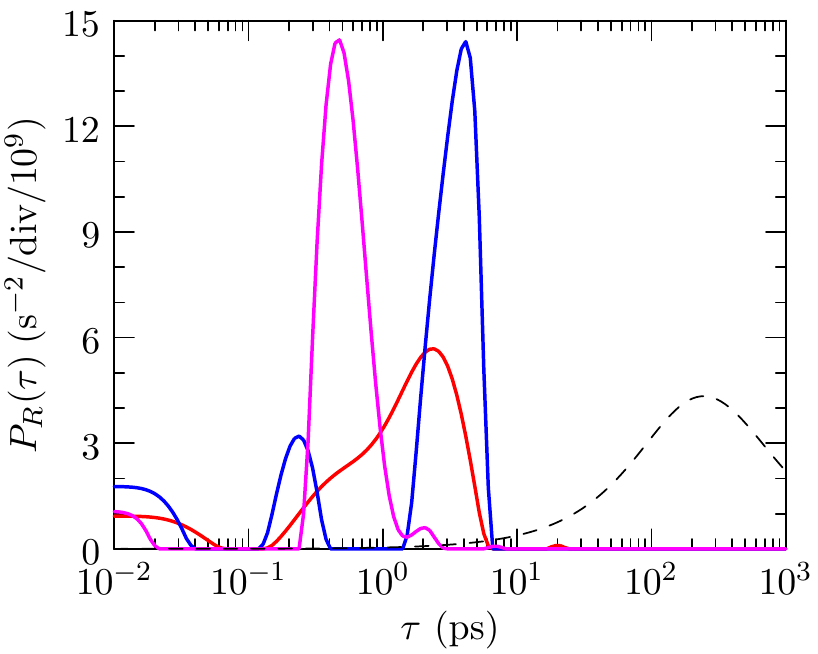}
\caption{\underline{Top}: Normalized autocorrelation for intra-molecular relaxation. Symbols are the simulation data (Fig.~\ref{fg:Intrabulk}); for clarity only every 5$^{th}$ point for $n$-heptane and water and every point for neopentane are shown. The curves are fits based on Eq.~\ref{eq:ILT1}. The inset highlights the short time behavior. \underline{Bottom}: Probability distribution of modes in Eq.~\ref{eq:ILT1}. As explained in the text, the dashed curve is the ``BPP frequency filter" defined at 400 MHz.}\label{fg:intraILT}
 \end{figure}

Figure~\ref{fg:intraILT} shows the results of the ILT analysis for the total intra-molecular autocorrelation function. As already emphasized, the mono-exponential Bloembergen-Purcell-Pound model is not adequate in capturing the $G_{R}$ data. However, the ILT procedure is able to accurately describe the available data. 

Figure~\ref{fg:intraILT} (bottom panel) shows the probability distribution of the molecular modes that help us recapitulate $G_R$ using Eq.~\ref{eq:ILT1}. A molecular interpretation of the modes remains an outstanding question that we are currently investigating, but we know by comparison of autocorrelations of rigid and flexible molecules that, for example, the mode around $\tau \approx 10^{-2}$~ps for $n$-heptane arises due to the motion of the terminal methyl group. The dashed line Fig.~\ref{fg:intraILT} (bottom panel) corresponds to the ``BPP frequency filter'' defined in Eq.~\ref{eq:JwfromPt}, at $f_0$ = 400 MHz (as an example). In other words, the components of $P_{R}(\tau)$ contributing to $T_1$ at $f_0 = 400$ MHz are weighted by the BPP frequency filter curve, which peaks at $\omega_0 \tau = 0.615$. The advantage of knowing $P_{R}(\tau)$ is then readily apparent, for we can clearly see that the components in $P_{R}(\tau)$ at $\tau \approx 10^{-2}$ ps do not contribute much to $T_1$ at $f_0 = 400$ MHz (as an example).

\begin{figure}[h!]
\includegraphics[scale=0.9]{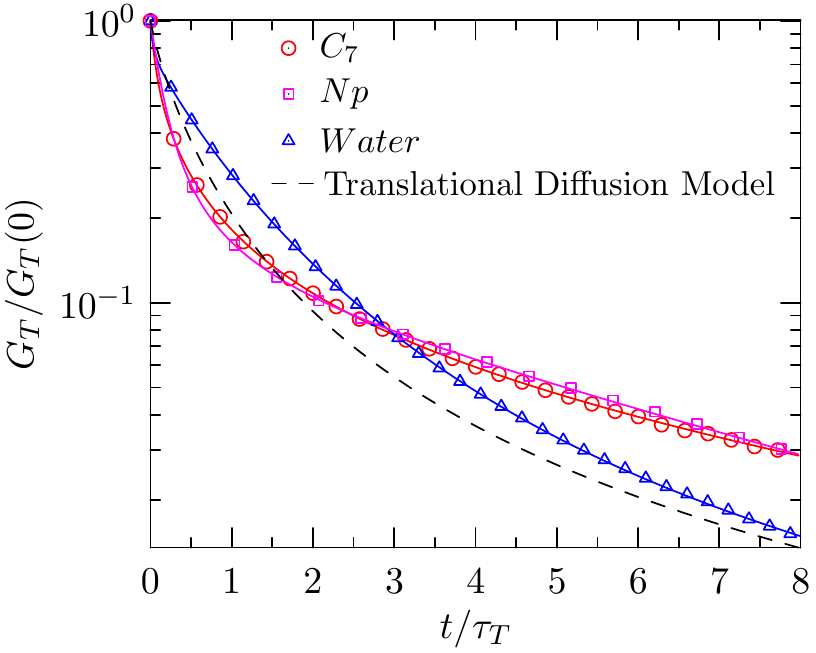}\\
\includegraphics[scale=0.9]{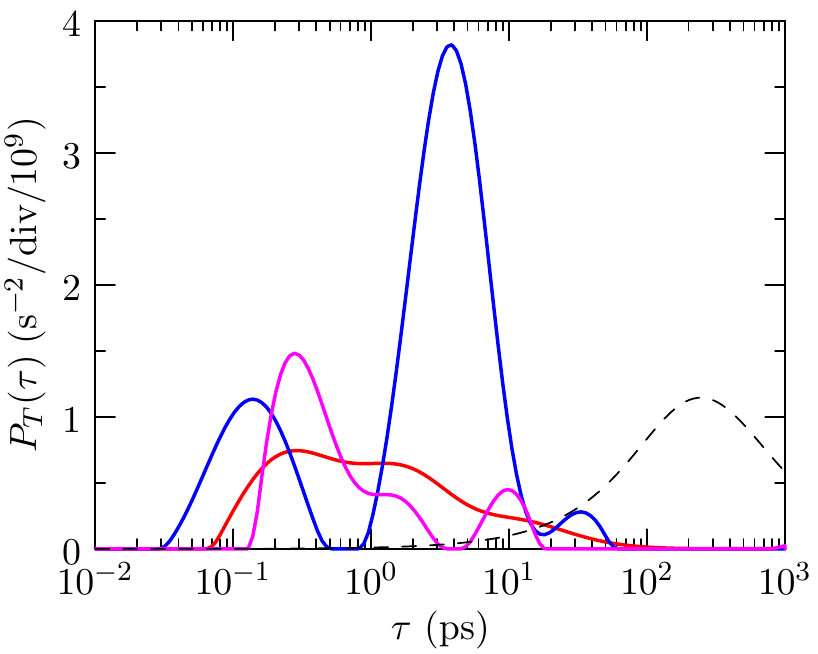}
\caption{\underline{Top}: Normalized autocorrelation for inter-molecular relaxation. Symbols are the simulation data (Fig.~\ref{fg:InterBulk}); for clarity only every 10$^{th}$ point is shown. The curves are fits based on Eq.~\ref{eq:ILT1}. \underline{Bottom}: Probability distribution of modes in Eq.~\ref{eq:ILT1}. Rest as in Fig.~\ref{fg:intraILT}.}\label{fg:interILT}
 \end{figure}

Figure~\ref{fg:interILT} shows the results of the ILT analysis for the total inter-molecular autocorrelation function. Clearly, $G_{T}(t)$ for inter-molecular $^1$H-$^1$H spin-pairs also lends itself to a multi-exponential description, Eq.~\ref{eq:ILT1}. Compared to the intra-molecular distribution, the ILT of $G_{T}$ reveals a broader distribution $P_{T}(t)$ in correlation times. 

The analysis above shows that using the ILT to predict $P_{R,T}(\tau)$ does not require a specific model for $G_{R,T}(t)$ such as the mono-exponential BPP model (Eq. \ref{eq:SumExp}) or the use of stretched exponents  \cite{singer:prb2020}. The justification for the multi-exponential (i.e. stretched) nature of $G_{R}(t)$ in Eq. \ref{eq:ILT1} comes from Woessner's theoretical treatment of NMR relaxation of intra-molecular $^1$H-$^1$H spin-pairs \cite{woessner:jcp1962,woessner:jcp1965}. Woessner's theories show that anisotropic rotation \cite{woessner:jcp1962} and internal motions \cite{woessner:jcp1965} of a spin-pair gives rise to a multi-exponential decay in $G_{R}(t)$. Furthermore, the larger the anisotropy and internal motion of the spin-pair, the more exponential decays are required to describe $G_{R}(t)$, implying a broader distribution in correlation times $P_{R}(\tau)$. 

 In the case of viscous fluids \cite{singer:jpcb2020} and fluids under confinement \cite{parambathu:jpcb2020}, we have shown that this approach yields good agreement with $T_1$ and $T_2$ measurements, even at low frequencies $f_0 = 2.3$ MHz. Furthermore, as discussed in \cite{singer:jmr2017,parambathu:jpcb2020}, $P_{R,T}(\tau)$ yields insights into the contribution of collective molecular modes in the observed autocorrelation, which can prove useful for interpreting $T_1$ and $T_2$ dispersion.

\subsubsection{Limitations of the ILT procedure}

There are two limitations in using  Eq. \ref{eq:ILT1}. The first limitation is that the ILT will not work in cases where a power-law decay in autocorrelation function $G_{R,T}(t) \propto t^{-\beta}$ ($1/4 \lesssim \beta \lesssim 1 $) is observed \cite{chavez:macro2011,chavez:macro2011b}. This is expected 
for viscoelastic materials or elastomers, such as highly entangled polymer melts. This limitation stems from the fact that $P_{R,T}(\tau) \propto  \tau^{-(1+\beta)}/\Gamma(\beta)$ (where $\Gamma $ is the gamma function, and $\beta>0$) diverges as $\tau \rightarrow 0$ for power-law decay of the form $G_{R,T}(t)\propto t^{-\beta}$.

The second limitation of ILT is the lack of sensitivity at long correlation times, or short relaxation times equivalently. There are two free parameter in the ILT, the regularization parameter $\alpha$ (Eq.~\ref{eq:reg}), and the maximum correlation time $\tau_{max}$ in $P_{R,T}(\tau)$. The maximum value of $\tau_{max}$ in $P_{R,T}(\tau)$ is chosen to be a factor 10 larger than the longest acquisition time $ t_{max} $, i.e. $\tau_{max} = 10\, t_{max}$, in accordance with  \cite{venkataramanan:ieee2002,song:jmr2002} and references within. This leads to inaccuracies in $P_{R,T}(\tau)$ if there are contributions with $\tau > \tau_{max}$. Consequently, the minimum $T_{1,2}$ in the fast-motion (i.e. low-frequency) regime is given by:
\begin{align}
\frac{1}{T_{1,2,min}} \simeq \frac{10}{3} \Delta\omega^2 \tau_{max} \simeq \frac{10}{3} \Delta\omega^2 10\, t_{max}   \label{eq:T1min}
\end{align}
where $\Delta\omega^2$ is the second moment (i.e. strength) of the interaction \cite{singer:jmr2017}. Using typical values of $\Delta\omega_R/2\pi \simeq $ 20 kHz for intra-molecular interactions \cite{singer:jmr2017}, and given the current computational limitations of $t_{max} = 10^3$ ps \cite{singer:jpcb2020,parambathu:jpcb2020}, leads to a minimum relaxation time of $T_{1,2,min} \simeq 2 $ ms. In other words, the ILT cannot determine relaxation times shorter than $T_{1,2} \lesssim T_{1,2,min} \simeq 2 $ ms. This limitation in the ILT did not hinder the prediction of $T_{1,2}$ down to $T_{1,2} \simeq 20 $ ms in the case of heptane confined in a polymer matrix \cite{parambathu:jpcb2020} or viscous polymers \cite{singer:jpcb2020}, where good agreement was found with measurements in both cases. Nevertheless, efforts are currently underway to extend $t_{max} $ beyond $t_{max} = 10^3$ ps.

The other free parameter in ILT is the regularization parameter $\alpha$. Given that the residual between the $G_{R,T}(t)$ data and the ILT fit is not dominated by Gaussian noise, we fix the regularization parameter to $\alpha =$ 10$^{-1}$ based on agreement with measurements \cite{singer:jcp2018,singer:jcp2018b,singer:jpcb2020,parambathu:jpcb2020}. We note however, that predictions of $T_{1,2}$ dispersion from ILT are much more sensitive to $\tau_{max}$ in Eq. \ref{eq:T1min} than to $\alpha$, therefore the selection of $\alpha$ is less critical than $\tau_{max}$. 

\section{Conclusions}
The Bloembergen-Purcell-Pound (BPP) model for intra-molecular NMR dipole-dipole relaxation and the Torrey, Hwang and Freed model for inter-molecular relaxation are important structures in the effort to use NMR to probe the the behavior of liquids. However, there is a danger in using the ideas in these models beyond the limits of their applicability. Specifically, the BPP model predicts a mono-exponential decay in the intra-molecular autocorrelation function between spin-pairs, which we confirm is the case between spin-pairs on the high-symmetry molecule neopentane. However, for molecules of lower symmetry such as $n$-heptane and water, the autocorrelation function between distinct spin pairs within the same molecule evince a stretched-exponential decay, implying a distribution in rotational correlation times. Likewise, modeling the inter-molecular autocorrelation function between spin pairs for a given initial separation using a mono-exponential is not accurate. Such assumptions will cause inaccurate predictions of the NMR relaxation dispersion (i.e. frequency dependence) in fluids.

Our work to date \cite{singer:jmr2017,singer:jcp2018,asthagiri:seg2018,parambathu:jpcb2020,singer:jpcb2020,singer:jcp2018b} shows that provided we have reasonable forcefields, MD simulation techniques can predict NMR relaxation in good agreement with measurements, without any adjustable parameters in the interpretation of the simulation data. Further, expanding the auto-correlation function in terms of molecular modes, where the molecular modes do have an exponential relaxation behavior, can account for the stretched exponential decay of the autocorrelation function and can be used to determine the NMR relaxation dispersion, without having to assume a model of molecular motion. The probability distribution of the molecular modes also have the potential to
enhance the molecular-scale interpretation of NMR relaxation times. Besides their utility in enhancing the interpretation of NMR relaxation experiments, the ideas presented here may prove useful
in efforts to interpret molecular relaxation behavior in other contexts in physical chemistry and in reconstructing autocorrelations from sparse data. 

\begin{acknowledgement}

We thank Edmund J. Fordham for discussions on using MaxEnt in NMR and for helpful comments on Laplace transforms. DA thanks Lawrence Pratt for many rewarding discussions on MaxEnt methods in general. We thank Arjun Valiya Parambathu for helpful comments on the manuscript. We thank Chevron Energy Technology Company, the Rice University Consortium on Processes in Porous Media, and the American Chemical Society Petroleum Research Fund (No. ACS PRF  58859-ND6) for financial support. 
We gratefully acknowledge the National Energy Research Scientific Computing Center, which is supported by the Office of Science of the U.S. Department of Energy (No.\ DE-AC02-05CH11231) and the Texas Advanced Computing Center (TACC) at The University of Texas at Austin for high-performance computer time and support. 
\end{acknowledgement} 

\newpage

\providecommand{\latin}[1]{#1}
\makeatletter
\providecommand{\doi}
  {\begingroup\let\do\@makeother\dospecials
  \catcode`\{=1 \catcode`\}=2 \doi@aux}
\providecommand{\doi@aux}[1]{\endgroup\texttt{#1}}
\makeatother
\providecommand*\mcitethebibliography{\thebibliography}
\csname @ifundefined\endcsname{endmcitethebibliography}
  {\let\endmcitethebibliography\endthebibliography}{}

 \end{document}